# Common-path generation of stable cylindrical perfect vector vortex beams with arbitrary order

Arabinda Mandal, Satyajit Maji and Maruthi M. Brundavanam*

*Department of Physics, Indian Institute of Technology Kharagpur, West Bengal-721302, India*
*Corresponding author: bmmanoj@phy.iitkgp.ac.in*

**Abstract:** A Highly flexible and efficient method of generating stable radially and Azimuthally polarized perfect optical vortex beams and all higher order cylindrical vector vortex beams, is proposed and demonstrated. The method is the most convenient implementation of the superposition of two orthogonally circularly polarized optical vortex beams of arbitrary integer topological charges. By simply controlling the relative amplitude and phase between vertical and horizontal polarization component of an input perfect vortex beam on a phase sensitive Spatial Light Modulator (SLM) and using a common-path interferometry, all arbitrary order Poincaré beams are prepared. Calibration curve of relative phase vs gray-scale of the phase function on the SLM is drawn to facilitate the determination of required phase-offset using pre-calculated gray-scale color-map. Generated beams states of polarization are uniformly distributed throughout the beam cross-section. The interference pattern of the two beam in a projected linear polarization state also gives the cylindrically symmetric petal beams with a control over the number and orientation of petals.

**Introduction:**

In recent times cylindrical vector vortex beams (CVVB) namely radially polarized (RP), azimuthally polarized (AP) and higher order Poincaré (HOP) beams is found to be of immense importance in many applications. A large number of instruments like photo lithography, confocal microscopy, optical tweezers etc. require sharper focused spot sizes. RP beam has stronger longitudinal and non-propagating electric field at the focus than any other polarized light beam and has the smallest focus spot (beyond the diffraction limit) and center along optic axis [1-2]. AP beam has donut shape intensity profile at focal plane due to polarization symmetry and its magnetic field is radially symmetric like electric field in RP. It is proposed that AP beam can be used in lighter magnetic particle trapping [3]. HOP sphere describes higher order basis states of polarization (SOP) of generalized vector vortex beams where basis is more general compared to first order Poincaré sphere [4]. Though both are Bloch sphere and both has orthogonal basis states, but HOP sphere's orthogonal basis states contains both spin and orbital angular momentum (OAM) of photon. Positive and negative index of HOP beam is decided by the sign of OAM. CVBBs have received extensive attention in optical trapping, super resolution imaging and quantum informatics [5]. Detection of longitudinal and angular speed of cooperative target by vector vortex beam has been reported recently [6].

The generation of vector beam can be mainly divided into two categories; one by internal control where SOP is controlled by specially designed resonator cavity with specially designed optical elements for each vector beam [7-10] and another by external control; by wavefront reconstruction using different optical elements [11-20]. Both processes suffer from intrinsic drawbacks. Control inside the cavity is a tedious process, as for a particular SOP a particular cavity is required. Also some extra special element like conical Brewster prism [9], anisotropic crystal [10] are used which can give only particular SOP beam. Whereas, for some of the later cases suffer from high instability as they use interferometric setup [18, 20], and some cases use special gratings [11, 12]. So their quality of construction defines efficiency and SOP of vector beams. It is also not easily controllable, and not easily convertible from one SOP to others. Particular elements give particular vector beam, such as lateral displacement beam splitter can give petals kind of vector beams, produce only RP and AP beams [19]. Recently, generation of HOP beam with perfect vortex by controlling incident polarization and using q-plate and circular dammann gratings has been reported [21]. Although vector vortex beam generation by inline interferometry using spatial light modulator (SLM) has been reported earlier but the setup used provides a low conversion efficiency, required very precise alignments and adjustments [22].

So generation of CVVBs (RP, AP and all HOP beams) with highly stable, easily controllable and convertible, uniform SOP with a single setup by using commonly used optical elements is in high demand. Here we present a convenient and easy approach to generate all order cylindrical perfect vector vortex beams (CPVVB) according to requirement from a single setup without using any special optical element, by only controlling the relative phase between horizontal and vertical polarization component of electric field on SLM. We have used the polarization selective phase modulation property of a nematic liquid crystal based phase only SLM. This is the first time to the best our knowledge that generation of CPVVB is demonstrated using such a method. Generated beams are very stable, SOP is uniform, easily controllable and convertible from one SOP to other by simply changing the spiral phase functions on SLM. Our generator composed of few commonly used optical elements such as Half Wave Plate (HWP), Quarter Wave Plates (QWP), polarizers, and SLMs.

**Theory:**

Perfect optical vortex (POV) is the optical beam which has a vortex phase, a narrow ring intensity structure and the radius of the ring does not depend upon the topological charge (TC) and with the highest gradient of the electric field on its boundary. Mathematically POV is the Fourier transform of Bessel-Gauss (BG) beam and its mathematical expression in cylindrical co-ordinates is given as [23]

$$E(r,\theta) = \frac{k}{f} i^{(m-1)} e^{il\theta} \int_0^\infty J_m(K_r\rho) J_m(\frac{K_r\rho}{f})\rho d\rho \ . \tag{1}$$

$J_m$ is an $m^{th}$ order Bessel function of first kind, m is the TC, f is the focal length of lens used to perform the Fourier transform. In order to generate POV, amplitude SLM is used with transmission function

$$H(x,y) = \frac{1}{2\pi} mod[m\tan^{-1}\left(\frac{y}{x}\right) - \frac{2\pi}{L} R\cos\left(\tan^{-1}\left(\frac{y}{x}\right)\right) + 2\pi \frac{R}{R_0}] \tag{2}$$

here R is the radial variable and $R_0$ is the scale factor of POV beams.

It is very well known fact that RP beam can be produced by the superposition of left circularly (LC) polarized vortex of TC +1, with right circularly (RC) polarized vortex of TC -1 [22]. Similarly, AP beam can be produced by the superposition of LC polarized vortex of TC +1, with RC polarized vortex of TC -1 with an extra phase difference π [22]. Whereas other combinations of TC and phase gives arbitrary higher order or negative indexed vector vortex beams.

For a CVVB the polarization pattern can be simply expressed as,

$$\Phi = q\theta + \alpha \qquad (3)$$

where $\Phi$ represents the orientation of the linear polarization, q is the Poincaré-Hopf index, θ is the azimuthal angle and α is the off-set angle at θ=0. For q=1, α=0 and π/2 represents radial polarization and azimuthal polarization, respectively.

**Experimental Method and Results:**

To demonstrate our technique, we implement the setup illustrated schematically in Fig. 1. The two SLMs employed are silicon based computer controlled nematic liquid crystal micro-displays, LC-R 1080 (amplitude only) and Pluto (phase only) from Holoeye.

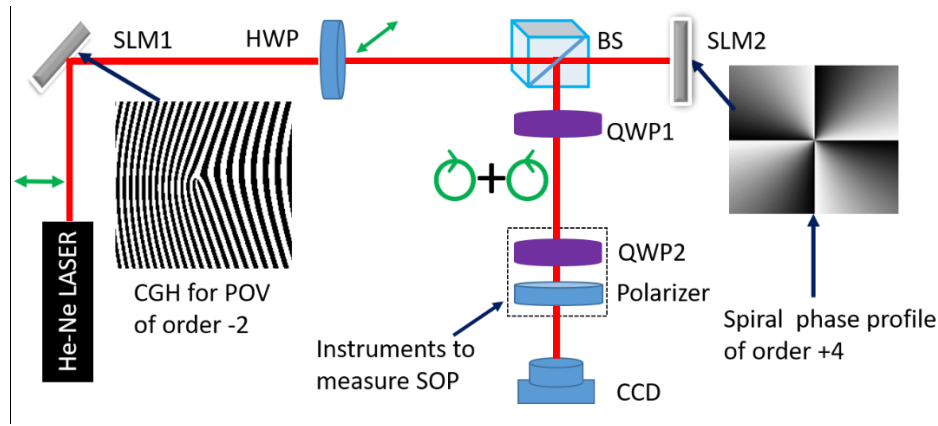

Fig1. (color online) Schematic of experimental setup for formation of vector vortex beams. HWP (QWP): Half (Quarter) Wave Plate, BS: Beam Splitter, SLM: Spatial Light Modulator. Green arrow indicates the polarization state.

The phase only SLM2 is pre-calibrated for a maximum phase change of 2π at 633 nm wavelength for a maximum gray-scale of 256 (0-255). When the phase only SLM displays a typical phase function through the calibrated gray-scale color-map, only the polarization component of the incoming light along a particular direction (in this case, along the longer dimension of the rectangular display, which is kept horizontal) is affected by the phase function. So, a spiral phase pattern displayed on the SLM would generate a symmetric OV beam when the input polarization is horizontal (H). The other polarization (i.e. vertical (V)) component of the incoming light will not see the phase function on the SLM and thus do not acquire the phase. Thus for an arbitrarily polarized incident beam of light, the output from the SLM will be a superposition of the phase modulated H-polarized component and an unmodulated V-polarized component. The H to V polarized component intensity ratio ($I_H/I_V$) can be very easily controlled by just changing the angle of the input linear polarization.

In order to generate POV of TC -1 we use amplitude only SLM (Holoeye LC-R 1080) by putting computer generated hologram (CGH) designed using equation 2 (m=-1). In reality any vortex (perfect or normal) beam generation method is applicable at this point. The only requirement is that a symmetric vortex beam of integer order needs to be generated. After passing through HWP the POV is converted to diagonally polarized light which has equal projection of horizontal and vertical components. CGH on SLM2 is elementary spiral phase function of order 2. So horizontal polarization component of POV beam gets modulated by SLM2 (phase only SLM) and acquires TC of +1 (-1+2=1) and vertical polarization component remains with TC -1. QWP1 kept at 45 degrees with respect to horizontal transform the horizontal and vertical polarization component to RC and LC, respectively. QWP2 and polarizer is used for Stokes parameter measurement. In order to generate RP beam, we change gray-scale offset of spiral phase function until it looks like Fig 2.a, when polarizer pass axis and fast axis of QWP2 are horizontally oriented.

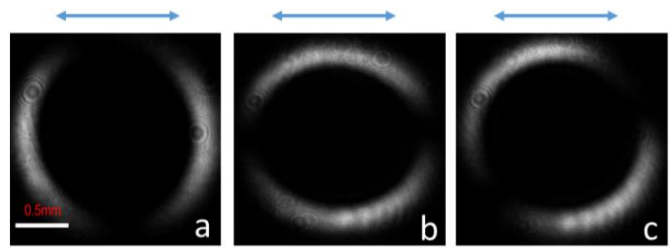

Fig2. Intensity pattern of (a) RP and (b) AP (c) Spiral polarization when analysed by a horizontal oriented polarizer. Blue arrow indicates orientation of polarizer.

To ensure that the generated beam is positive indexed beam (RP, AP and other HOP of positive index) one has to rotate the polarizer by removing QWP2. If the rotation of polarizer and intensity pattern are same than the beam is positive indexed beam, otherwise it is negative indexed beam (Hedgehog beam and HOP of negative indexed). To convert negative index to positive index, one needs to rotate QWP1 by 90 degrees (i.e. from -45 degree to +45 degree). SOP of the beam is measured using Stokes polarimetry method [24], and the SOP is shown in Fig 3.a, which is RP beam. For generating AP beam, the gray-level offset of spiral phase function needs to be increased until it looks like Fig 2.b by putting QWP2 and polarizer at 0 degree. SOP of generated AP beam is shown in Fig. 3.b. Beam with spiral SOP can be generated by adjusting the gray-scale offset of optical elementary function in between RP and AP so that it looks like Fig. 2.c when axes of polarizer is at horizontal. By putting QWP1 at -45 degree and measuring the Stokes parameter we got the SOP Fig. 3.c which is negative indexed beam with Hedgehog polarization.

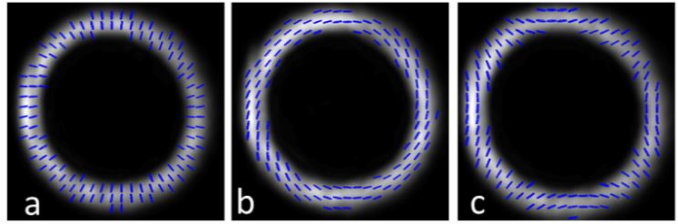

Fig.3. (color online) Experimental SOP distribution of (a) RP (b) AP (both of index +1) and (c) Hedgehog (index -1)

Increase in the offset of gray-scale color-map basically changes the relative phase between the horizontally polarized and vertically polarized component of the POV beam. Fig.4 shows how the relative phase between horizontal and vertical component of electric field is changing by increasing the gray scale offset on SLM2. Slope and intercept of this calibration curve are found to be 0.025 radian/gray scale value and -0.15 rad. So any required relative phase can be generated by simply changing the gray-scale offset.

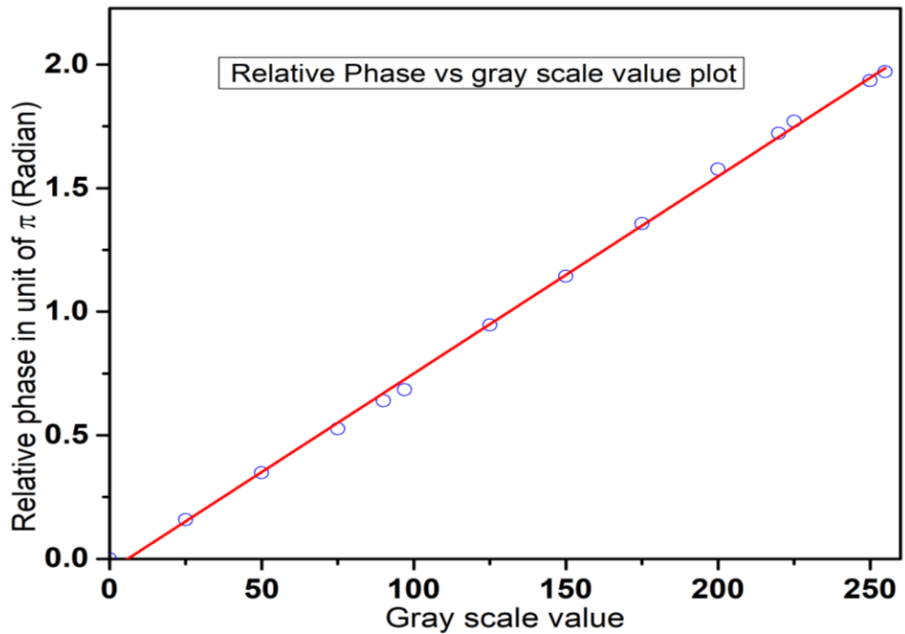

Fig4. (color online) Calibration curve of imposed relative phase on H-polarized component with respect to V-polarized component due to different gray scale offset shown on the phase SLM.

For generation of HOP beam of order m, we generate POV of higher order (-m) using equation 2 and set SLM2 of higher order (2m) spiral phase function and adjust the relative phase between horizontal and vertical component by changing the gray level on SLM2. Fig.5 shows the symmetric petal shaped intensity patterns of generated HOP beams of indices 0 to +10 except 1 (unit indexed beam is already shown in Fig 2.a). The number of azimuthal dark lines is twice of the index of CPVVBs. This kind of optical beams has found utility in simultaneous optical trapping of multiple micro-particles, laser processing and structure formation on materials [25].

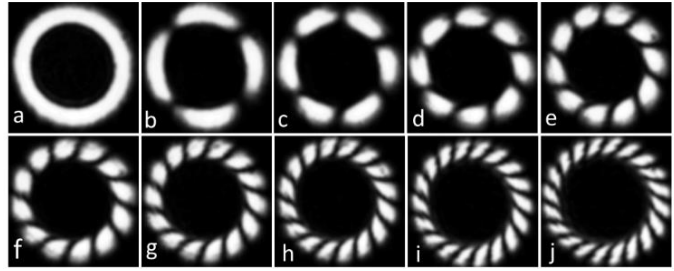

Fig.5. Circular symmetric intensity pattern of vector vortex beam of different index when analysed through a horizontal polarizer. (a-j) shows CVVBs of index 0 to 10 except 1 (already shown in Fig 2.a).

SOPs of the HOP beams are measured by same method as previous measurement. Fig 6.a – 6.c show the SOP distribution of negative indexed HOP beams of order -2 to -4 and Fig. 6.d - 6.f show that of positive indexed HOP beams of order 2 to 4, respectively.

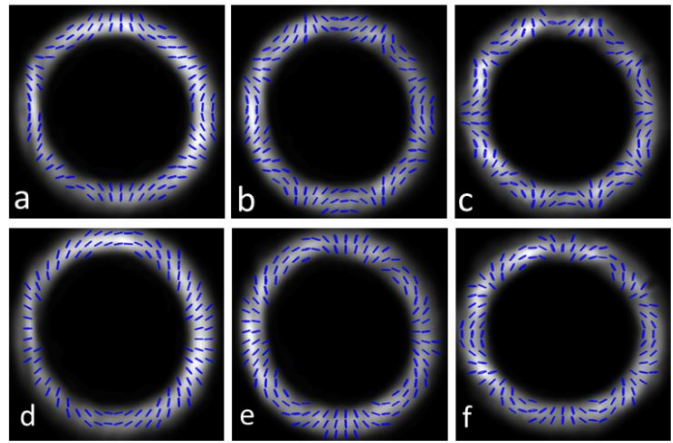

Fig 6. (Color online) Experimental polarization distribution of negative indexed HOP beams of indices (a) -2, (b) -3, (c) -4 and positive indexed HOP beams of indices (d) 2, (e) 3, (f) 4.

## Conclusion:

 We have proposed and demonstrated a simple, highly stable and easily controllable process to generate both positive and negative indexed unit and all higher ordered cylindrical perfect vector vortex beams by only controlling the relative phase by SLM by simply utilizing the fact that phase only SLM modulates a particular linearly polarized light (in our case horizontal polarized). Here conversion from one SOP to other of same order require a small change in gray-scale offset following a measured calibration curve to get useful perfect vector vortex beams. Our experimental setup composed of few commonly used optical elements and does not require any special optical elements like many previous reported works. From SLM calibration curve we can calculate the required relative phase and change the gray level offset of the phase function on SLM accordingly to get different HOP beam. Although our method is very easy, flexible and efficient but use of two SLMs increases the cost of setup. But the point to be noted is that one can use any commonly used process to generate OV/POV instead of using amplitude SLM. One can also use the phase SLM itself to generate the POV by splitting the active area of the SLM window in two part while designing the CGHs thus reducing the number of SLM employed from two to one.